%% file: main.tex
\documentclass[sigconf]{acmart}

\usepackage{circledsteps}
\usepackage{tablefootnote}
\usepackage{enumitem}
\usepackage[export]{adjustbox}
\usepackage{listings}

\definecolor{mygreen}{rgb}{0,0.6,0}
\definecolor{mygray}{rgb}{0.5,0.5,0.5}
\definecolor{mymauve}{rgb}{0.58,0,0.82}
\definecolor{terminalbgcolor}{HTML}{330033}
\definecolor{terminalrulecolor}{HTML}{000099}

\lstdefinestyle{terminal}{
	backgroundcolor=\color{terminalbgcolor},
	basicstyle=\tiny\color{white}\ttfamily\selectfont,
	breakatwhitespace=true,  
	breaklines=true,
	captionpos=b,
	commentstyle=\color{mygreen},
	deletekeywords={...},
	escapeinside={\%*}{*)},
	extendedchars=true,
	frame=single,
	keepspaces=true,
	keywordstyle=\color{blue},
	morekeywords={*,...},
	numbers=none,
	numbersep=5pt,
        framerule=1pt,
	numberstyle=\color{mygray}\tiny\selectfont,
	rulecolor=\color{terminalrulecolor},
	showspaces=false,
	showstringspaces=false,
	showtabs=false,
	stepnumber=2,
	stringstyle=\color{mymauve},
	tabsize=2,
        xleftmargin=5pt,
        xrightmargin=5pt
}

\lstdefinestyle{html}{
	backgroundcolor=\color{white},
	basicstyle=\color{black}\ttfamily\selectfont,
	breakatwhitespace=true,  
	breaklines=true,
	captionpos=b,
	deletekeywords={...},
        mathescape=true,
        escapechar=\%
	extendedchars=true,
	frame=single,
	keepspaces=true,
	morekeywords={*,...},
	numbers=none,
	numbersep=5pt,
        framerule=1pt,
	numberstyle=\color{mygray}\tiny\selectfont,
	rulecolor=\color{black},
	showspaces=false,
	showstringspaces=false,
	showtabs=false,
	stepnumber=2,
	stringstyle=\color{mymauve},
	tabsize=2,
        xleftmargin=5pt,
        xrightmargin=5pt
}

\AtBeginDocument{%
  \providecommand\BibTeX{{%
    \normalfont B\kern-0.5em{\scshape i\kern-0.25em b}\kern-0.8em\TeX}}}

\copyrightyear{2024}
\acmYear{2024}
\setcopyright{rightsretained}
\acmConference[ICSE-Companion '24]{2024 IEEE/ACM 46th International Conference on Software Engineering: Companion Proceedings}{April 14--20, 2024}{Lisbon, Portugal}
\acmBooktitle{2024 IEEE/ACM 46th International Conference on Software Engineering: Companion Proceedings (ICSE-Companion '24), April 14--20, 2024, Lisbon, Portugal}\acmDOI{10.1145/3639478.3640022}
\acmISBN{979-8-4007-0502-1/24/04}

\begin{document}

\title{CATMA: Conformance Analysis Tool \\For Microservice Applications}


\author{Clinton Cao}
\affiliation{%
  \institution{Delft University of Technology}
  \country{The Netherlands}
}
\author{Simon Schneider}
\affiliation{%
  \institution{Hamburg University of Technology}
  \country{Germany}
}
\author{Nicolás E. Díaz Ferreyra}
\affiliation{%
  \institution{Hamburg University of Technology}
  \country{Germany}
}
\author{Sicco Verwer}
\affiliation{%
  \institution{Delft University of Technology}
  \country{The Netherlands}
}
\author{Annibale Panichella}
\affiliation{%
  \institution{Delft University of Technology}
  \country{The Netherlands}
}
\author{Riccardo Scandariato}
\affiliation{%
  \institution{Hamburg University of Technology}
  \country{Germany}
}

\renewcommand{\shortauthors}{Cao and Schneider, et al.}

\begin{abstract}
The microservice architecture allows developers to divide the core functionality of their software system into multiple smaller services. However, this architectural style also makes it harder for them to debug and assess whether the system's deployment conforms to its implementation. We present CATMA, an automated tool that detects non-conformances between the system's deployment and implementation. It automatically visualizes and generates potential interpretations for the detected discrepancies. Our evaluation of CATMA shows promising results in terms of performance and providing useful insights. CATMA is available at \url{https://cyber-analytics.nl/catma.github.io/}, and a demonstration video is available at \url{https://youtu.be/WKP1hG-TDKc}.
\end{abstract}

\ccsdesc[500]{Software and its engineering~Software testing and debugging}
\ccsdesc[500]{Software and its engineering~Automated static analysis}
\ccsdesc[500]{Software and its engineering~Dynamic analysis}

\keywords{microservices, static analysis, dynamic analysis, software testing, empirical software engineering}


\maketitle

\input{introduction}
\input{running_example}
\input{catma}

\input{tool_validation}
\input{related_work}
\input{conclusion_and_future_work}

\begin{acks}
\noindent 
We thank the colleagues who participated in CATMA's pilot study. This work was partly funded by the European Union's Horizon 2020 program under grant agreement No. 952647 (AssureMOSS).
\end{acks}

\balance
\bibliographystyle{ACM-Reference-Format}
\bibliography{references}

\end{document}

%% file: introduction.tex
\section{Introduction}\label{sec:intro}
Software systems following the microservice architectural paradigm have their core functionality split into multiple smaller components.
These microservices (or just \textit{services}) of a microservice application (MSA) communicate via lightweight communication protocols such as REST APIs or message brokers.
The services of an MSA can be developed, maintained, and deployed independently, paving the way for an increasing trend in the adoption of this architectural style.
Despite these benefits, MSAs pose a challenge in gaining a comprehensive overview due to their inherently decoupled and distributed nature. 
Consequently, debugging faults is a time-consuming process because the localization of the root cause is challenging.
According to studies, developers usually take several days to debug and find the cause of a fault~\cite{Zhou2018_fault_analysis,Lenarduzzi_Panichella_serverless}. 
Many approaches for the automatic extraction of architectural representations of MSA have been proposed~\cite{alshuqayran18_misar, Granchelli2017_microart, Ma2019_vmamv, soldani21_mtosca}, thus addressing the challenge of gaining an overview of the applications' architecture.
Some approaches combine static and dynamic analysis to build the architectural models.
Also, multiple fault localization techniques for MSAs have been proposed~\cite{Zhou2019_latent,Guo2020_graph}, which use dynamic analysis to identify faults and pinpoint the root cause in code.
However, to the best of our knowledge, no work compares the results from static and dynamic analysis rather than merging them.
Moreover, none of the existing fault localization approaches offer explainability in the form of possible interpretations for the faults.

In this paper, we present CATMA, a novel tool designed to analyze and compare statically and dynamically obtained architectural models. CATMA autonomously identifies potential non-conformances between these models, generating easily accessible visualizations for users and providing concise interpretations. 
These interpretations reduce the number of lines in source code that users need to scrutinize when investigating a non-conformance.
We tested CATMA on four open-source MSAs and conducted a preliminary usability study with two participants.
The results indicate that the tool effectively supports developers during the localization and debugging of non-conformances, demonstrating its usefulness and potential in the debugging landscape for microservices.



%% file: running_example.tex
\section{Running Example}\label{sec:example}
The software engineering team of ZYX Inc. is working on their new web application for selling tech products. They embrace the microservice architectural style as this allows them to split up into smaller groups and work independently on the core functionalities of their application. Each member follows the best practices of software engineering; using static analysis to detect faults and testing each functionality before its deployment. After finishing the development, they deploy the application to test it out. To their surprise, they notice that the monitoring service does not receive any metrics data. They are unsure of the cause of this discrepancy since a static analysis tool correctly detects the line of code that implements the transmission of metrics data and does not raise any warnings. They spend several days analyzing different log files, but have no luck in finding the underlying cause. They scratch their heads and start wondering whether there is a tool that provides: 
\begin{itemize}[leftmargin=*]
    \item the detection of discrepancies between the implementation and deployment of MSAs,
    \item a high-level overview of such discrepancies, and
    \item descriptions of the potential root causes.
\end{itemize}

%% file: catma.tex
\section{CATMA}\label{sec:catma}

\textbf{Worfklow}.
\begin{figure}
    \centering
    \includegraphics[width=0.97\columnwidth]{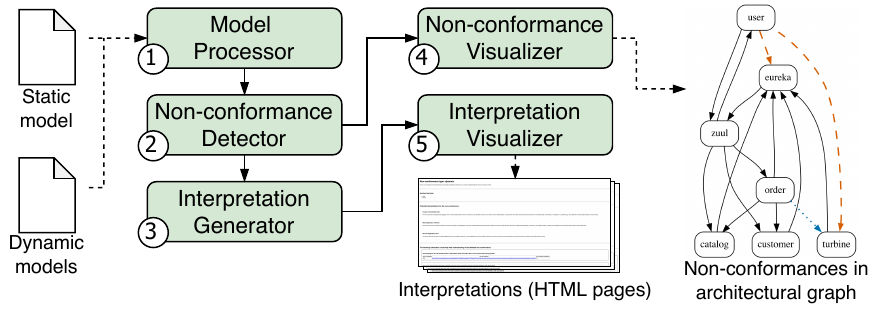}
    \vspace{-2mm}
    \caption{CATMA's workflow. Input models are processed(\Circled{1}) and non-conformances between them detected (\Circled{2}) and visualized (\Circled{4}). Each non-conformance is visualized (\Circled{5}) and possible interpretations for it are generated (\Circled{3}).}
    \label{fig:tool_workflow}
    \vspace{-2mm}
\end{figure}
Figure~\ref{fig:tool_workflow} depicts CATMA's workflow. First, the \textit{Model Processor}~\Circled{1} reads the input models (static and dynamic) to extract architectural components. The obtained data is passed on to the \textit{Non-conformance Detector}~\Circled{2}, which checks whether there are any non-conformances (discrepancies) between static and dynamic models. If a non-conformance is detected, it is forwarded to both the \textit{Interpretation Generator}~\Circled{3} and the \textit{Non-conformance Visualizer}~\Circled{4}. The latter (\Circled{4}) collects all detected non-conformances and generates a visualization of the system's architecture that shows the non-conformances. The former (\Circled{3}) generates a set of possible interpretations for each detected non-conformance, which describe potential causes. These interpretations are forwarded to the \textit{Interpretation Visualizer}~\Circled{5}, which generates HTML pages that visualize the interpretations. CATMA is designed to be modular. Each component can be replaced or expanded to fit the user's needs. 
The tool is invoked via the command line (see Listing~\ref{lst:commandline}).



\begin{lstlisting}[caption={Command-line invocation of CATMA.}, label={lst:commandline}, style=terminal]
$ ~/Doc/Git/CATMA python3 CATMA.py \
--static_model_path data/ewolff_microservice/ewolff_microservice_static_model.json \
--dynamic_models_path data/ewolff_microservice/dynamic_models/ \ 
--output_path ./output/
Reading configuration file...
Processing static model...
Processing dynamic model...
Detecting non-conformances: 100% |=====|13/13 [00:00<00:00, 83245.73it/s]
Detecting non-conformances: 100% |=====|13/13 [00:00<00:00, 152733.76it/s]
Detected 2 static non-conformances and 1 dynamic non-conformance 
between implementation and deployment of the system!
Generating non-conformance interpretations...
Generating non-conformance visualizations...
Generating interpretation visualizations...
\end{lstlisting}

\textbf{Detecting Non-conformances}.
%
As static models, CATMA accepts dataflow diagrams (DFDs) like the ones introduced by Schneider and Scandariato~\cite{schneider_code2dfd23}. These DFDs are automatically extracted from source code and configuration files by searching for relevant keywords and using them as evidence to build relationships between services.
As dynamic models, state machines inferred from HTTP events logs are expected. They are created using a similar model-inference approach as presented by Cao et al.~\cite{cao_learning_state_machines22}. The approach first extracts logs from a Kubernetes cluster using Packetbeat. It then utilizes Flexfringe~\cite{Verwer17_flexfringe} to generate behavioral traces and learns a state machine from these traces. 
The \textit{Model Processor} extracts services and connections between them from both input models. 
They represent the application's architecture and are used to detect non-conformances.
In the DFD, nodes and edges depict the services and information flows between them, respectively. We can, therefore, directly extract the nodes and edges. 
In a state machine, services and their corresponding relations are represented differently; each transition in a state machine indicates which services in the system have communicated with each other.
Thus, nodes and edges are extracted from the transitions of the state machines.
The \textit{Model Processor} creates a set of nodes and edges for both input models, where edges are represented as \textit{“service X $\rightarrow$ service Y”} and denote the communication relationship between the two services.




Non-conformances are detected by identifying differences between the sets of nodes and edges.
The \textit{Non-conformance Detector} iterates through the sets and checks for each item whether it exists in both corresponding sets. 
We define \textit{static non-conformances} as nodes or edges missing from the static model (compared to the dynamic model) and \textit{dynamic non-conformances} as those missing from the dynamic model.
Each item is tagged according to this comparison, i.e., indicating whether it is present in both, only the static, or only the dynamic model.
The tagged sets of nodes and edges are passed to components \Circled{3} and \Circled{4}.

The \textit{Non-conformance Visualizer} is responsible for creating a graphical representation of any detected non-conformances.
It generates a PlantUML file (see plantuml.com) which presents the nodes and edges as a graph and where a coloring scheme highlights any found non-conformances.
Model items observed in both models are colored black, items only observed in the static model (dynamic non-conformances) are colored blue, and items only observed in the dynamic model (static non-conformances) are colored orange. In addition, dynamic and static non-conformances are visually distinguished by means of dotted and dashed lines, respectively.

\textbf{Interpreting Non-conformances}.
CATMA generates a set of possible interpretations for each detected non-conformance. These interpretations are visualized in an HTML page by the \textit{Interpretation Visualizer}. The HTML page helps users analyze the potential causes of non-conformances. 
CATMA presents a specific set of interpretations for both types of non-conformance. The generated HTML pages contain (1) the type, definition, and involved services of the non-conformance, (2) the set of possible interpretations, and (3) additional details that support the understanding of the non-conformance. In the following, (2) and (3) are described further.



\begin{figure}
    \centering
    \includegraphics[width=\columnwidth, frame]{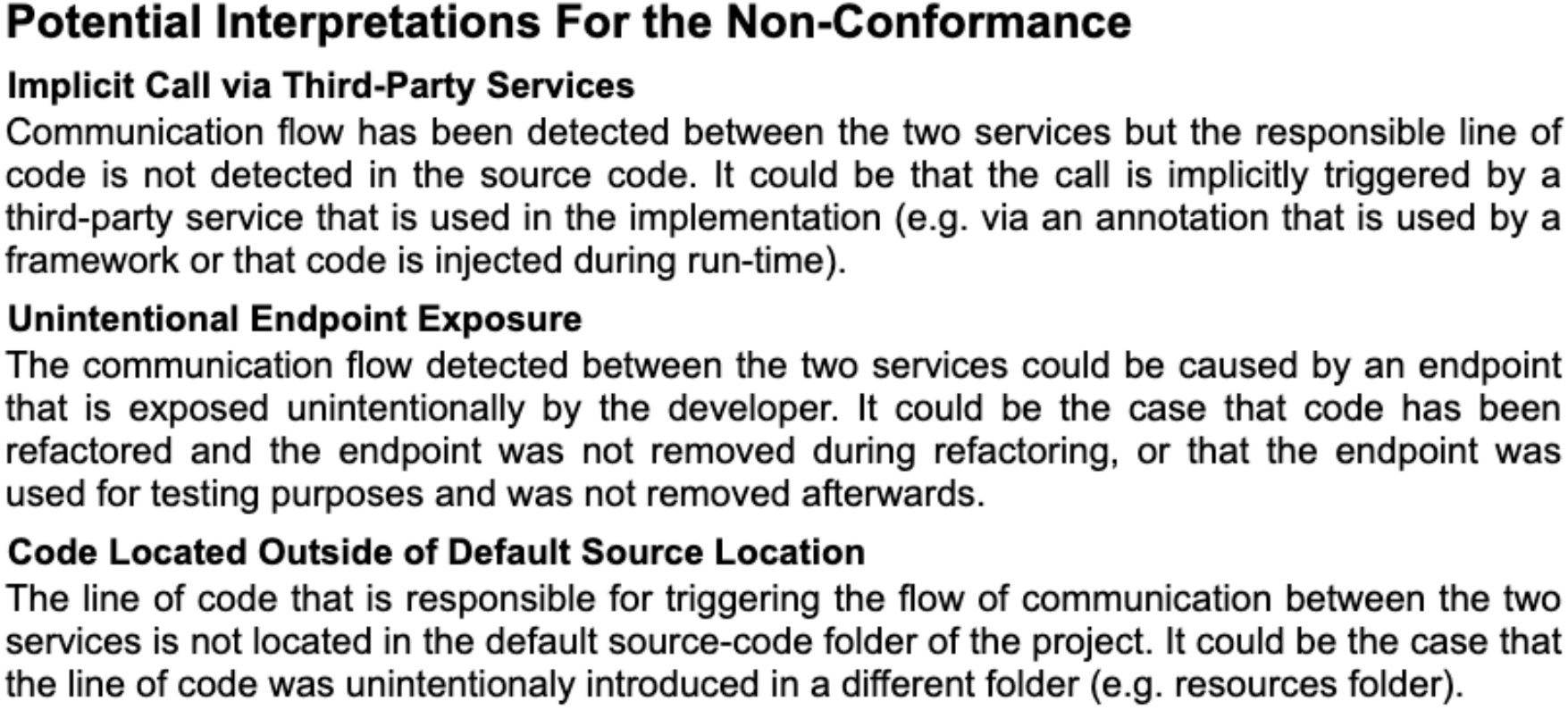}
    \vspace{-6mm}
    \caption{Example set of textual interpretations. }
    \label{fig:example_interpretation_set}
    \vspace{-7mm}
\end{figure}

\textit{Providing Interpretations of Non-Conformances}.
A set of high-level textual interpretations is provided, which describe possible underlying causes of the detected non-conformances.
The interpretations are meant to serve as possible starting points to debug found non-conformances. 
Currently, the generation is based solely on the type of non-conformance, i.e., whether it is static or dynamic.
We formulated a text describing possible interpretations for both types of non-conformances, and the corresponding one is presented to the user.
As the basis for these interpretations, we collected known causes of non-conformances from the literature (e.g., ~\cite{Hannousse2021_securing,waseem_on_the_nature_2021}). These causes range from standard programming errors made in software development to common causes for issues encountered by developers of MSAs. As an example, misconfiguration of services is a common cause of dynamic non-conformances in MSAs. When services are not properly configured, they become undiscoverable by other services, leading to missing expected runtime behaviors. CATMA uses this information as a basis for the generation of one interpretation for a dynamic non-conformance.
For the collection, we disregarded non-conformances rooted in hardware-related issues, e.g., due to non-deterministic behavior because of multi-threading or similar effects. 
Figure~\ref{fig:example_interpretation_set} presents the set of textual interpretations that are provided for a static non-conformance.

Our future work will predominantly focus on this part of the tool, specifically on implementing a more intelligent generation of applicable interpretations.
In this regard, we will analyze indicators for each cause of non-conformances.
These indicators will then be used to decide whether a cause is plausible or not for a given non-conformance.
This will lead to the generation of a tailored set of possible interpretations for each found non-conformance.
The already carried-out analysis of the related literature provides the basis for this future work.

\begin{figure}
    \centering
    \includegraphics[width=\columnwidth, frame]
    {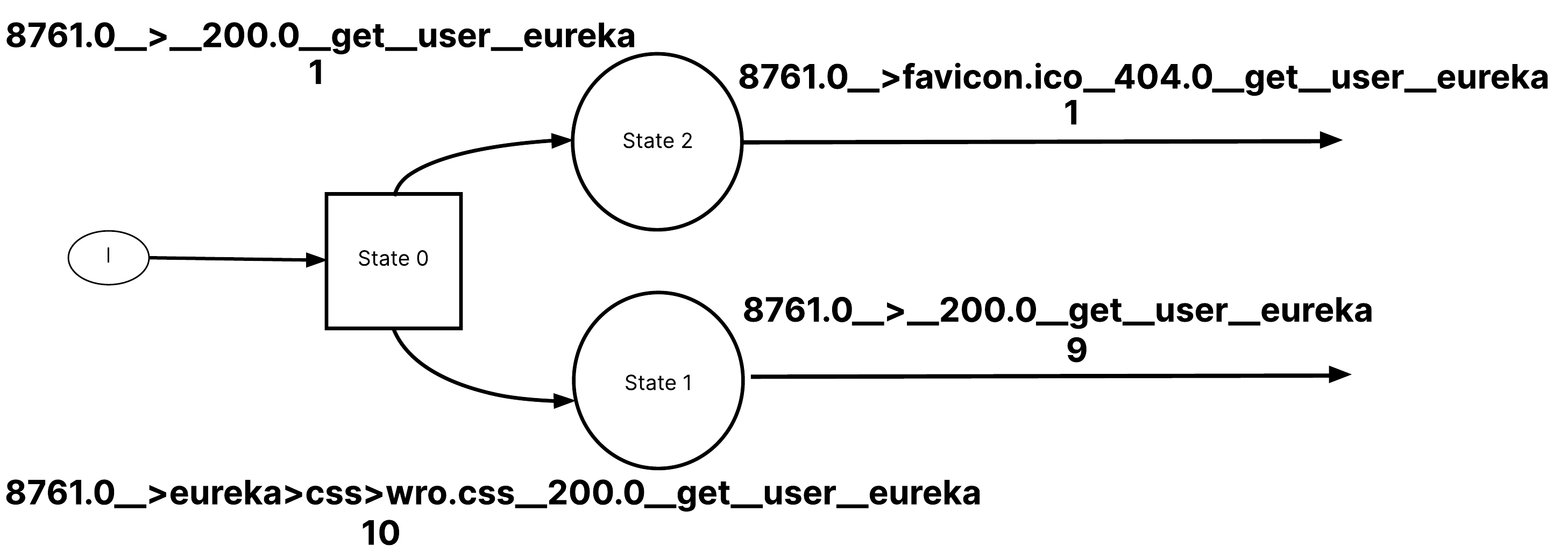}
    \vspace{-6mm}
    \caption{Part of state machine showing unexpected behavior.}
    \label{fig:example_state_machine_link}
    \vspace{-3mm}
\end{figure}

\begin{figure}
    \centering
    \includegraphics[width=0.96\columnwidth, frame]{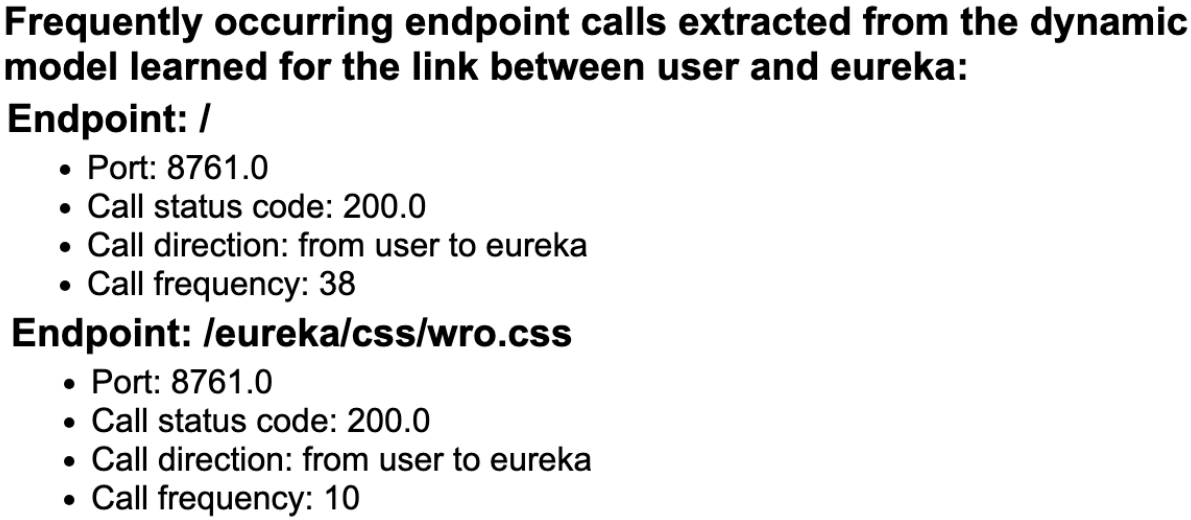}
    \vspace{-3mm}
    \caption{Most frequent calls for unexpected behavior.}
    \label{fig:example_readable_calls}
    \vspace{-3mm}
\end{figure}

\begin{figure}
    \centering
    \includegraphics[width=\columnwidth]{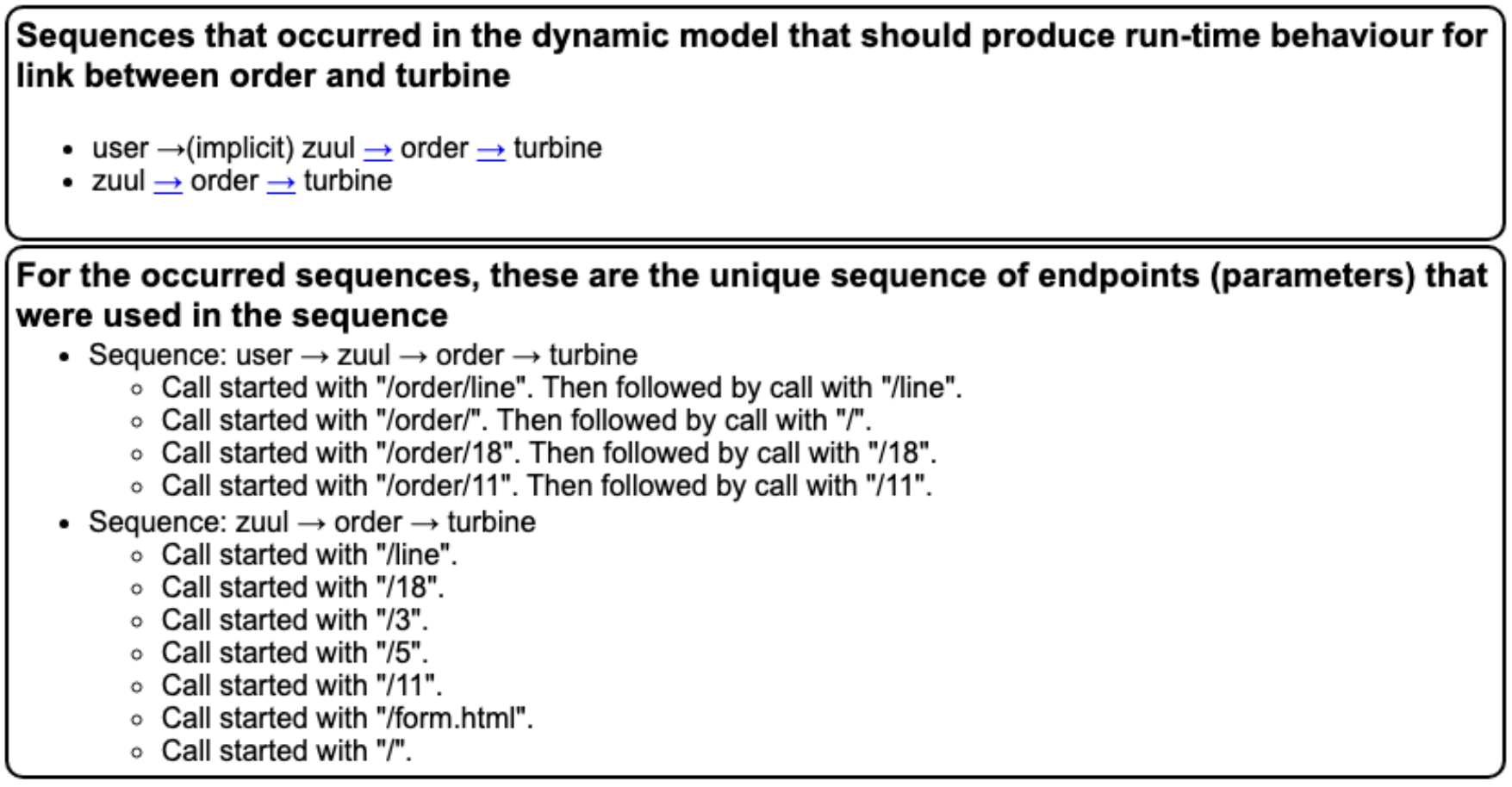}
    \vspace{-7mm}
    \caption{Example details for dynamic non-conformance.}
    \label{fig:example_dynamic_details}
    \vspace{-1mm}
\end{figure}

\textit{Additional details}.
The generated HTML page also presents additional details that could aid the user with the understanding of the detected non-conformance. In the case of static non-conformances, a state machine is visualized that depicts the unexpected sequential communication behavior detected between the involved services. The most frequently occurring calls between the involved services are presented in a human-readable format right after the state machine model. This insight can be used to understand why such calls were made between the involved services. Figures~\ref{fig:example_state_machine_link} and~\ref{fig:example_readable_calls} show an example of a state machine and the most frequent calls, respectively. In the case of a dynamic non-conformance, we instead leverage the traceability information contained in the static model to point to the code that shows the expected behavior. 
Specifically, the page presents (1) the line of code responsible for triggering the missing runtime event (i.e., the line of code that should have been executed), (2) the sequence of events that should trigger the missing runtime event, and (3) human-readable call details extracted for the previous point. Figure~\ref{fig:example_dynamic_details} provides a snapshot of this set of details.
Furthermore, the state machines learned for each involved service are presented on the HTML page.

%% file: tool_validation.tex
\section{Tool Evaluation}\label{sec:tool_evaluation}

\textbf{Performance Analysis}.
We evaluated CATMA's performance in terms of time to detect non-conformances in MSA.
For this, we selected 4 DFDs of open-source MSAs from the dataset created by Schneider et al. \cite{schneider_microSecEnD23}, deployed these MSAs, and created state machine models for them.
Then, we ran CATMA on the obtained models and measured the time of the analysis.
Table~\ref{table:runtime_msa} presents the time for analyzing the 4 selected MSAs (averaged over 10 executions per MSA).
%
%
This evaluation allows us to quantify the benefits of utilizing CATMA compared to manual analysis. 
\begin{table}
\centering
\vspace{-1mm}
\caption{CATMA's performance statistics on multiple MSAs}\label{table:runtime_msa}
\vspace{-2mm}
\resizebox{\columnwidth}{!}{
\begin{tabular}{l|c|c|c|c}
\toprule
\multicolumn{1}{c|}{Name}                                                & \#LOC & \# Services & \begin{tabular}[c]{@{}c@{}}\# Detected \\ Non-Conformances  \\(static / dynamic)\end{tabular} & \begin{tabular}[c]{@{}c@{}}Avg. \\ Runtime (seconds)\end{tabular} \\ \midrule
\textit{Springboot-Microservice}\tablefootnote{https://github.com/shabbirdwd53/springboot-microservice}           & 879   & 9           & 0 / 16                                                                      & 4.3                                                              \\
\textit{Microservice Sample}\tablefootnote{https://github.com/ewolff/microservice}                            & 3117  & 7           & 2 / 1                                                                       & 3.0                                                              \\
\textit{Spring PetClinic}\tablefootnote{https://github.com/spring-petclinic/spring-petclinic-microservices} & 3990  & 12          & 1 / 26                                                                      & 78.6                                                            \\
\textit{Piggy Metrics}\tablefootnote{https://github.com/sqshq/piggymetrics}                             & 9977  & 17          & 3 / 11                                                                      & 53.9                                                            \\ \bottomrule
\end{tabular}
}
\vspace{-4mm}
\end{table}

The data clearly demonstrates that CATMA substantially accelerates the analysis process. While developers often invest days in resolving issues (as reported in the study conducted by Zhou et al.~\cite{Zhou2018_fault_analysis}), our tool accomplishes the same task in a matter of minutes. Thus, CATMA can substantially reduce the time spent on debugging issues, offering a valuable resource for developers.

\textbf{Pilot Study}.
We conducted a small-scale pilot study to investigate CATMA's usefulness. We report an initial assessment of this pilot study based on a think-aloud interview setup with two participants. The participants were recruited from the lab of one of the authors (both with a computer science background) and have no relation to the work done for CATMA. The participants got an introduction to MSAs and were allowed to interact with CATMA before the start of the interview. During the interview, we asked several questions that would provide us insights on what are the most useful elements presented in the output generated by CATMA. A complete transcript of the interview can be found on our Figshare page~\cite{cao_2023_appendix_catma}. The following points summarize the most useful elements from CATMA's output: (1) the model-based visualization that shows where non-conformances are detected, (2) the set of possible interpretations providing the potential causes for the corresponding non-conformance, (3) the ability to jump from the dynamic model (state machine) back to the source code, (4) static non-conformances provide insights on the security implication of the system, and (5) the type of the non-conformances: static non-conformances provide insights on the security implication.

\textbf{Correctness of Dynamic Models}.
As the state machines approximate the provided log data, it is helpful to understand the trade-off between the correctness and the size of the model as it could influence the detection of non-conformances; a small state machine generalizes too much and introduces inaccuracies, a large state machine captures all possible behavior but might be hard to understand and process. To evaluate this aspect, we use a technique similar to the one proposed by Walkinshaw et al.~\cite{Walkinshaw2016_inferring}. Table~\ref{table:model_correctness} presents the average results computed from a 10-fold cross-validation experiment. As expected, smaller state machines introduce more inaccuracies, leading to lower balanced accuracy scores. This suggests that smaller state machines do lead to more inaccuracies in the detection of non-conformances. Furthermore, the accuracy scores appear to plateau as the state machine grows in size. This suggests that considerably larger state machines do not perform significantly better in the detection of non-conformances and selecting the largest possible model for the detection is redundant. Learning a moderate-sized state machine from input data should provide reliable performance for detecting non-conformances.


\begin{table}
\centering
\small
\caption{Trade-off between the size and correctness.}
\label{table:model_correctness}
\vspace{-2mm}
\resizebox{\columnwidth}{!}{
\begin{tabular}{c|c|c|c|c}
\toprule
Avg. \# Edges & Avg. \# Nodes & Avg. Recall & Avg. Specificity & Avg. Balanced Accuracy \\ \midrule
1       & 1             & 0.0         & 1.0              & 0.5                    \\
127     & 99            & 0.368       & 0.998            & 0.683                  \\
790     & 646           & 0.904       & 0.990            & 0.947                  \\
1982    & 1843          & 0.920       & 0.986            & 0.953                  \\
4714    & 4570          & 1.0         & 0.978            & 0.989                  \\ \bottomrule
\end{tabular}
}
\vspace{-4mm}
\end{table}

%% file: related_work.tex
\section{Related work}\label{sec:related_work}

Several studies have demonstrated that architectural software representations can assist developers during manual system analysis activities~\cite{Arisholm06_impact_uml, Gravino15_code_comprehension_uml, Schneider24_DFDs_empirical_experiment}.
To automate such processes, several approaches in the related literature combine static and dynamic analysis for architecture reconstruction of MSAs.
\textit{MicroArt} presented by Granchelli et al.\cite{Granchelli2017_microart}, \textit{MiSAR} presented by Alshuqayran et al.~\cite{alshuqayran18_misar}, and $\mu$\textit{TOSCA} presented by Soldani et al.~\cite{soldani21_mtosca} all extract the list of microservices statically by parsing deployment files.
Connections between them are detected dynamically by leveraging service discovery services that exist in the analyzed applications or by injecting different monitoring tools.
\textit{VMAWV} presented by Ma et al.~\cite{Ma2019_vmamv} instead queries existing service discovery services to retrieve the list of services and uses static analysis to detect connections. While these approaches combine static and dynamic analysis, none of them compare complete architectural models obtained via the two techniques.
Since our approach performs this comparison to identify non-conformances, we believe it to be novel in this regard.

The approach \textit{DOMICO} by Zhong et al.~\cite{zhong23_domico} also detects non-conformances between system representations of different stages in the development process, however, they compare the intended design (UML) against the actual implementation (static model). This approach is partly based on the approach introduced by Murphy et al.~\cite{Murphy01_software}. The proposed approach by Murphy et al. computes a reflexion model by finding differences between the architectural model extracted from the source code and the mental (architectural) model constructed by a system developer. Both approaches detect non-conformances between design and implementation, whereas CATMA detects non-conformances between the system's implementation and deployment.

%% file: conclusion_and_future_work.tex
\section{Conclusion \& Future Work}\label{sec:conclusion}
We present CATMA, a tool for automatically conducting conformance analysis of MSAs. 
It detects possible non-conformances by computing differences between a statically and a dynamically obtained architectural model of the MSA.
Found non-conformances are visualized in an easily accessible way.
Further, a set of possible interpretations is generated, showing the non-conformances' potential causes.
In a preliminary evaluation, CATMA showed promising results in terms of performance as well as usability. 
In our evaluation, CATMA identified a non-conformance in an open-source MSA on GitHub. 
A misconfiguration in the Hystrix ~\footnote{https://github.com/Netflix/Hystrix} monitoring dashboard prevented stream data from being visualized as intended in the implementation.
This is a good example of a non-conformance between the intended and observed behaviors of the MSA.
We notified the developers and our fix was accepted~\footnote{https://github.com/ewolff/microservice/pull/30}.
Hence, CATMA has already shown its first --albeit small-- impact on MSA.




As future work, we will extend CATMA with a more intelligent technique for selecting suitable interpretations for found non-conformances.
Further, the approach would benefit from additional validation activities concerning its usefulness and possible enhancements.
We plan a user study with developers in which they identify non-conformances with the help of CATMA.  
Finally, we will investigate the feasibility of using other types of models as input and the detection capabilities of other non-conformances.

%% file: main.bbl

\begin{thebibliography}{21}


\ifx \showCODEN    \undefined \def \showCODEN     #1{\unskip}     \fi
\ifx \showDOI      \undefined \def \showDOI       #1{#1}\fi
\ifx \showISBNx    \undefined \def \showISBNx     #1{\unskip}     \fi
\ifx \showISBNxiii \undefined \def \showISBNxiii  #1{\unskip}     \fi
\ifx \showISSN     \undefined \def \showISSN      #1{\unskip}     \fi
\ifx \showLCCN     \undefined \def \showLCCN      #1{\unskip}     \fi
\ifx \shownote     \undefined \def \shownote      #1{#1}          \fi
\ifx \showarticletitle \undefined \def \showarticletitle #1{#1}   \fi
\ifx \showURL      \undefined \def \showURL       {\relax}        \fi
\providecommand\bibfield[2]{#2}
\providecommand\bibinfo[2]{#2}
\providecommand\natexlab[1]{#1}
\providecommand\showeprint[2][]{arXiv:#2}

\bibitem[Alshuqayran et~al\mbox{.}(2018)]%
        {alshuqayran18_misar}
\bibfield{author}{\bibinfo{person}{Nuha Alshuqayran}, \bibinfo{person}{Nour
  Ali}, {and} \bibinfo{person}{Roger Evans}.} \bibinfo{year}{2018}\natexlab{}.
\newblock \showarticletitle{Towards Micro Service Architecture Recovery: An
  Empirical Study}. In \bibinfo{booktitle}{\emph{2018 IEEE International
  Conference on Software Architecture (ICSA)}}. \bibinfo{pages}{47--4709}.
\newblock
\urldef\tempurl%
\url{https://doi.org/10.1109/ICSA.2018.00014}
\showDOI{\tempurl}


\bibitem[Arisholm et~al\mbox{.}(2006)]%
        {Arisholm06_impact_uml}
\bibfield{author}{\bibinfo{person}{E. Arisholm}, \bibinfo{person}{L.C. Briand},
  \bibinfo{person}{S.E. Hove}, {and} \bibinfo{person}{Y. Labiche}.}
  \bibinfo{year}{2006}\natexlab{}.
\newblock \showarticletitle{The impact of UML documentation on software
  maintenance: an experimental evaluation}.
\newblock \bibinfo{journal}{\emph{IEEE Transactions on Software Engineering}}
  \bibinfo{volume}{32}, \bibinfo{number}{6} (\bibinfo{year}{2006}),
  \bibinfo{pages}{365--381}.
\newblock
\urldef\tempurl%
\url{https://doi.org/10.1109/TSE.2006.59}
\showDOI{\tempurl}


\bibitem[Cao et~al\mbox{.}(2022)]%
        {cao_learning_state_machines22}
\bibfield{author}{\bibinfo{person}{Clinton Cao}, \bibinfo{person}{Agathe
  Blaise}, \bibinfo{person}{Sicco Verwer}, {and} \bibinfo{person}{Filippo
  Rebecchi}.} \bibinfo{year}{2022}\natexlab{}.
\newblock \showarticletitle{Learning State Machines to Monitor and Detect
  Anomalies on a Kubernetes Cluster}. In \bibinfo{booktitle}{\emph{Proceedings
  of the 17th International Conference on Availability, Reliability and
  Security}} (Vienna, Austria) \emph{(\bibinfo{series}{ARES '22})}.
  \bibinfo{publisher}{Association for Computing Machinery},
  \bibinfo{address}{New York, NY, USA}, Article \bibinfo{articleno}{117},
  \bibinfo{numpages}{9}~pages.
\newblock
\showISBNx{9781450396707}
\urldef\tempurl%
\url{https://doi.org/10.1145/3538969.3543810}
\showDOI{\tempurl}


\bibitem[Cao et~al\mbox{.}(2023)]%
        {cao_2023_appendix_catma}
\bibfield{author}{\bibinfo{person}{Clinton Cao}, \bibinfo{person}{Simon
  Schneider}, \bibinfo{person}{Nicolás E.~Ferreyra Diaz},
  \bibinfo{person}{Sicco Verwer}, \bibinfo{person}{A.~(Annibale) Panichella},
  {and} \bibinfo{person}{Riccardo Scandariato}.}
  \bibinfo{year}{2023}\natexlab{}.
\newblock \showarticletitle{{Appendix for 'CATMA: Conformance Analysis Tool for
  Microservice Applications'}}.
\newblock  (\bibinfo{date}{10} \bibinfo{year}{2023}).
\newblock
\urldef\tempurl%
\url{https://doi.org/10.6084/m9.figshare.23942214.v3}
\showDOI{\tempurl}


\bibitem[Granchelli et~al\mbox{.}(2017)]%
        {Granchelli2017_microart}
\bibfield{author}{\bibinfo{person}{Giona Granchelli}, \bibinfo{person}{Mario
  Cardarelli}, \bibinfo{person}{Paolo Di~Francesco}, \bibinfo{person}{Ivano
  Malavolta}, \bibinfo{person}{Ludovico Iovino}, {and} \bibinfo{person}{Amleto
  Di~Salle}.} \bibinfo{year}{2017}\natexlab{}.
\newblock \showarticletitle{MicroART: A Software Architecture Recovery Tool for
  Maintaining Microservice-Based Systems}. In \bibinfo{booktitle}{\emph{2017
  IEEE International Conference on Software Architecture Workshops (ICSAW)}}.
  \bibinfo{pages}{298--302}.
\newblock
\urldef\tempurl%
\url{https://doi.org/10.1109/ICSAW.2017.9}
\showDOI{\tempurl}


\bibitem[Gravino et~al\mbox{.}(2015)]%
        {Gravino15_code_comprehension_uml}
\bibfield{author}{\bibinfo{person}{Carmine Gravino}, \bibinfo{person}{Giuseppe
  Scanniello}, {and} \bibinfo{person}{Genoveffa Tortora}.}
  \bibinfo{year}{2015}\natexlab{}.
\newblock \showarticletitle{Source-code comprehension tasks supported by UML
  design models: Results from a controlled experiment and a differentiated
  replication}.
\newblock \bibinfo{journal}{\emph{Journal of Visual Languages \& Computing}}
  \bibinfo{volume}{28} (\bibinfo{year}{2015}), \bibinfo{pages}{23--38}.
\newblock
\showISSN{1045-926X}
\urldef\tempurl%
\url{https://doi.org/10.1016/j.jvlc.2014.12.004}
\showDOI{\tempurl}


\bibitem[Guo et~al\mbox{.}(2020)]%
        {Guo2020_graph}
\bibfield{author}{\bibinfo{person}{Xiaofeng Guo}, \bibinfo{person}{Xin Peng},
  \bibinfo{person}{Hanzhang Wang}, \bibinfo{person}{Wanxue Li},
  \bibinfo{person}{Huai Jiang}, \bibinfo{person}{Dan Ding},
  \bibinfo{person}{Tao Xie}, {and} \bibinfo{person}{Liangfei Su}.}
  \bibinfo{year}{2020}\natexlab{}.
\newblock \showarticletitle{Graph-Based Trace Analysis for Microservice
  Architecture Understanding and Problem Diagnosis}. In
  \bibinfo{booktitle}{\emph{Proceedings of the 28th ACM Joint Meeting on
  European Software Engineering Conference and Symposium on the Foundations of
  Software Engineering}} (Virtual Event, USA) \emph{(\bibinfo{series}{ESEC/FSE
  2020})}. \bibinfo{publisher}{Association for Computing Machinery},
  \bibinfo{address}{New York, NY, USA}, \bibinfo{pages}{1387–1397}.
\newblock
\showISBNx{9781450370431}
\urldef\tempurl%
\url{https://doi.org/10.1145/3368089.3417066}
\showDOI{\tempurl}


\bibitem[Hannousse and Yahiouche(2021)]%
        {Hannousse2021_securing}
\bibfield{author}{\bibinfo{person}{Abdelhakim Hannousse} {and}
  \bibinfo{person}{Salima Yahiouche}.} \bibinfo{year}{2021}\natexlab{}.
\newblock \showarticletitle{Securing microservices and microservice
  architectures: A systematic mapping study}.
\newblock \bibinfo{journal}{\emph{Computer Science Review}}
  \bibinfo{volume}{41} (\bibinfo{year}{2021}), \bibinfo{pages}{100415}.
\newblock
\showISSN{1574-0137}
\urldef\tempurl%
\url{https://doi.org/10.1016/j.cosrev.2021.100415}
\showDOI{\tempurl}


\bibitem[Lenarduzzi and Panichella(2021)]%
        {Lenarduzzi_Panichella_serverless}
\bibfield{author}{\bibinfo{person}{Valentina Lenarduzzi} {and}
  \bibinfo{person}{Annibale Panichella}.} \bibinfo{year}{2021}\natexlab{}.
\newblock \showarticletitle{Serverless Testing: Tool Vendors' and Experts'
  Points of View}.
\newblock \bibinfo{journal}{\emph{IEEE Software}} \bibinfo{volume}{38},
  \bibinfo{number}{1} (\bibinfo{year}{2021}), \bibinfo{pages}{54--60}.
\newblock
\urldef\tempurl%
\url{https://doi.org/10.1109/MS.2020.3030803}
\showDOI{\tempurl}


\bibitem[Ma et~al\mbox{.}(2019)]%
        {Ma2019_vmamv}
\bibfield{author}{\bibinfo{person}{Shang-Pin Ma}, \bibinfo{person}{I-Hsiu Liu},
  \bibinfo{person}{Chun-Yu Chen}, \bibinfo{person}{Jiun-Ting Lin}, {and}
  \bibinfo{person}{Nien-Lin Hsueh}.} \bibinfo{year}{2019}\natexlab{}.
\newblock \showarticletitle{Version-Based Microservice Analysis, Monitoring,
  and Visualization}. In \bibinfo{booktitle}{\emph{2019 26th Asia-Pacific
  Software Engineering Conference (APSEC)}}. \bibinfo{pages}{165--172}.
\newblock
\urldef\tempurl%
\url{https://doi.org/10.1109/APSEC48747.2019.00031}
\showDOI{\tempurl}


\bibitem[Murphy et~al\mbox{.}(2001)]%
        {Murphy01_software}
\bibfield{author}{\bibinfo{person}{G.C. Murphy}, \bibinfo{person}{D. Notkin},
  {and} \bibinfo{person}{K.J. Sullivan}.} \bibinfo{year}{2001}\natexlab{}.
\newblock \showarticletitle{Software reflexion models: bridging the gap between
  design and implementation}.
\newblock \bibinfo{journal}{\emph{IEEE Transactions on Software Engineering}}
  \bibinfo{volume}{27}, \bibinfo{number}{4} (\bibinfo{year}{2001}),
  \bibinfo{pages}{364--380}.
\newblock
\urldef\tempurl%
\url{https://doi.org/10.1109/32.917525}
\showDOI{\tempurl}


\bibitem[Schneider et~al\mbox{.}(2024)]%
        {Schneider24_DFDs_empirical_experiment}
\bibfield{author}{\bibinfo{person}{Simon Schneider},
  \bibinfo{person}{Nicolas~E. Diaz~Ferreyra}, \bibinfo{person}{Pierre-Jean
  Queval}, \bibinfo{person}{Georg Simhandl}, \bibinfo{person}{Uwe Zdun}, {and}
  \bibinfo{person}{Riccardo Scandariato}.} \bibinfo{year}{2024}\natexlab{}.
\newblock \showarticletitle{How Dataflow Diagrams Impact Software Security
  Analysis: an Empirical Experiment}. In \bibinfo{booktitle}{\emph{SANER}}.
\newblock


\bibitem[Schneider and Scandariato(2023)]%
        {schneider_code2dfd23}
\bibfield{author}{\bibinfo{person}{Simon Schneider} {and}
  \bibinfo{person}{Riccardo Scandariato}.} \bibinfo{year}{2023}\natexlab{}.
\newblock \showarticletitle{Automatic extraction of security-rich dataflow
  diagrams for microservice applications written in Java}.
\newblock \bibinfo{journal}{\emph{Journal of Systems and Software}}
  \bibinfo{volume}{202} (\bibinfo{year}{2023}), \bibinfo{pages}{111722}.
\newblock
\showISSN{0164-1212}
\urldef\tempurl%
\url{https://doi.org/10.1016/j.jss.2023.111722}
\showDOI{\tempurl}


\bibitem[Schneider et~al\mbox{.}(2023)]%
        {schneider_microSecEnD23}
\bibfield{author}{\bibinfo{person}{Simon Schneider}, \bibinfo{person}{Tufan
  Özen}, \bibinfo{person}{Michael Chen}, {and} \bibinfo{person}{Riccardo
  Scandariato}.} \bibinfo{year}{2023}\natexlab{}.
\newblock \showarticletitle{microSecEnD: A Dataset of Security-Enriched
  Dataflow Diagrams for Microservice Applications}. In
  \bibinfo{booktitle}{\emph{2023 IEEE/ACM 20th International Conference on
  Mining Software Repositories (MSR)}}. \bibinfo{pages}{125--129}.
\newblock
\urldef\tempurl%
\url{https://doi.org/10.1109/MSR59073.2023.00030}
\showDOI{\tempurl}


\bibitem[Soldani et~al\mbox{.}(2021)]%
        {soldani21_mtosca}
\bibfield{author}{\bibinfo{person}{Jacopo Soldani}, \bibinfo{person}{Giuseppe
  Muntoni}, \bibinfo{person}{Davide Neri}, {and} \bibinfo{person}{Antonio
  Brogi}.} \bibinfo{year}{2021}\natexlab{}.
\newblock \showarticletitle{The mTOSCA toolchain: Mining, analyzing, and
  refactoring microservice-based architectures}.
\newblock \bibinfo{journal}{\emph{Software: Practice and Experience}}
  \bibinfo{volume}{51}, \bibinfo{number}{7} (\bibinfo{year}{2021}),
  \bibinfo{pages}{1591--1621}.
\newblock
\urldef\tempurl%
\url{https://doi.org/10.1002/spe.2974}
\showDOI{\tempurl}


\bibitem[Verwer and Hammerschmidt(2017)]%
        {Verwer17_flexfringe}
\bibfield{author}{\bibinfo{person}{Sicco Verwer} {and}
  \bibinfo{person}{Christian~A. Hammerschmidt}.}
  \bibinfo{year}{2017}\natexlab{}.
\newblock \showarticletitle{flexfringe: A Passive Automaton Learning Package}.
  In \bibinfo{booktitle}{\emph{2017 IEEE International Conference on Software
  Maintenance and Evolution (ICSME)}}. \bibinfo{pages}{638--642}.
\newblock
\urldef\tempurl%
\url{https://doi.org/10.1109/ICSME.2017.58}
\showDOI{\tempurl}


\bibitem[Walkinshaw et~al\mbox{.}(2016)]%
        {Walkinshaw2016_inferring}
\bibfield{author}{\bibinfo{person}{Neil Walkinshaw}, \bibinfo{person}{Ramsay
  Taylor}, {and} \bibinfo{person}{John Derrick}.}
  \bibinfo{year}{2016}\natexlab{}.
\newblock \showarticletitle{Inferring Extended Finite State Machine Models from
  Software Executions}.
\newblock \bibinfo{journal}{\emph{Empirical Software Engineering}}
  \bibinfo{volume}{21}, \bibinfo{number}{3} (\bibinfo{date}{jun}
  \bibinfo{year}{2016}), \bibinfo{pages}{811–853}.
\newblock
\showISSN{1382-3256}
\urldef\tempurl%
\url{https://doi.org/10.1007/s10664-015-9367-7}
\showDOI{\tempurl}


\bibitem[Waseem et~al\mbox{.}(2021)]%
        {waseem_on_the_nature_2021}
\bibfield{author}{\bibinfo{person}{Muhammad Waseem}, \bibinfo{person}{Peng
  Liang}, \bibinfo{person}{Mojtaba Shahin}, \bibinfo{person}{Aakash Ahmad},
  {and} \bibinfo{person}{Ali~Rezaei Nassab}.} \bibinfo{year}{2021}\natexlab{}.
\newblock \showarticletitle{On the Nature of Issues in Five Open Source
  Microservices Systems: An Empirical Study}. In
  \bibinfo{booktitle}{\emph{Proceedings of the 25th International Conference on
  Evaluation and Assessment in Software Engineering}} (Trondheim, Norway)
  \emph{(\bibinfo{series}{EASE '21})}. \bibinfo{publisher}{Association for
  Computing Machinery}, \bibinfo{address}{New York, NY, USA},
  \bibinfo{pages}{201–210}.
\newblock
\showISBNx{9781450390538}
\urldef\tempurl%
\url{https://doi.org/10.1145/3463274.3463337}
\showDOI{\tempurl}


\bibitem[Zhong et~al\mbox{.}(2023)]%
        {zhong23_domico}
\bibfield{author}{\bibinfo{person}{Chenxing Zhong}, \bibinfo{person}{He Zhang},
  \bibinfo{person}{Huang Huang}, \bibinfo{person}{Zhikun Chen},
  \bibinfo{person}{Chao Li}, \bibinfo{person}{Xiaodong Liu}, {and}
  \bibinfo{person}{Shanshan Li}.} \bibinfo{year}{2023}\natexlab{}.
\newblock \showarticletitle{DOMICO: Checking conformance between domain models
  and implementations}.
\newblock \bibinfo{journal}{\emph{Software: Practice and Experience}}
  (\bibinfo{year}{2023}).
\newblock
\urldef\tempurl%
\url{https://doi.org/10.1002/spe.3272}
\showDOI{\tempurl}


\bibitem[Zhou et~al\mbox{.}(2018)]%
        {Zhou2018_fault_analysis}
\bibfield{author}{\bibinfo{person}{Xiang Zhou}, \bibinfo{person}{Xin Peng},
  \bibinfo{person}{Tao Xie}, \bibinfo{person}{Jun Sun}, \bibinfo{person}{Chao
  Ji}, \bibinfo{person}{Wenhai Li}, {and} \bibinfo{person}{Dan Ding}.}
  \bibinfo{year}{2018}\natexlab{}.
\newblock \showarticletitle{Fault analysis and debugging of microservice
  systems: Industrial survey, benchmark system, and empirical study}.
\newblock \bibinfo{journal}{\emph{IEEE Transactions on Software Engineering}}
  \bibinfo{volume}{47}, \bibinfo{number}{2} (\bibinfo{year}{2018}),
  \bibinfo{pages}{243--260}.
\newblock


\bibitem[Zhou et~al\mbox{.}(2019)]%
        {Zhou2019_latent}
\bibfield{author}{\bibinfo{person}{Xiang Zhou}, \bibinfo{person}{Xin Peng},
  \bibinfo{person}{Tao Xie}, \bibinfo{person}{Jun Sun}, \bibinfo{person}{Chao
  Ji}, \bibinfo{person}{Dewei Liu}, \bibinfo{person}{Qilin Xiang}, {and}
  \bibinfo{person}{Chuan He}.} \bibinfo{year}{2019}\natexlab{}.
\newblock \showarticletitle{Latent Error Prediction and Fault Localization for
  Microservice Applications by Learning from System Trace Logs}. In
  \bibinfo{booktitle}{\emph{Proceedings of the 2019 27th ACM Joint Meeting on
  European Software Engineering Conference and Symposium on the Foundations of
  Software Engineering}} (Tallinn, Estonia) \emph{(\bibinfo{series}{ESEC/FSE
  2019})}. \bibinfo{publisher}{Association for Computing Machinery},
  \bibinfo{address}{New York, NY, USA}, \bibinfo{pages}{683–694}.
\newblock
\showISBNx{9781450355728}
\urldef\tempurl%
\url{https://doi.org/10.1145/3338906.3338961}
\showDOI{\tempurl}


\end{thebibliography}
